\documentclass[conference]{IEEEtran}

\usepackage{cite}
\usepackage{booktabs}
\usepackage{graphicx}
\usepackage{amsmath,amssymb,amsfonts}
\usepackage{multirow}
\usepackage{array}
\usepackage{glossaries}
\newacronym{fmr}{FMR}{False Mapping Ratio}
\DeclareMathOperator*{\argmax}{argmax}

\newcommand{\capec}{CAPEC attack pattern}
\newcommand{\capecs}{CAPEC attack patterns}
\newcommand{\ics}{ATT\&CK ICS technique}
\newcommand{\icss}{ATT\&CK ICS techniques}

\usepackage{tikz}

\newcommand\submittedtext{%
  \footnotesize This work has been submitted to the IEEE for possible publication. Copyright may be transferred without notice, after which this version may no longer be accessible.}

\newcommand\submittednotice{%
\begin{tikzpicture}[remember picture,overlay]
\node[anchor=south,yshift=10pt] at (current page.south) {\fbox{\parbox{\dimexpr0.65\textwidth-\fboxsep-\fboxrule\relax}{\submittedtext}}};
\end{tikzpicture}%
}

\AtBeginDocument{
    \providecommand\BibTeX{{
    \normalfont B\kern-0.5em{\scshape i\kern-0.25em b}\kern-0.8em\TeX}
    }
}

\begin{document}

\title{Cyber Knowledge Completion Using Large Language Models}

\author{
    \IEEEauthorblockN{
        Braden K Webb, Sumit Purohit, Rounak Meyur
    }
    \IEEEauthorblockA{
        Pacific Northwest National Laboratory \\
        Richland, Washington 99352 \\
        \{braden.webb,sumit.purohit,rounak.meyur\}@pnnl.gov
    }
}
% \author{Braden K Webb, Sumit Purohit, Rounak Meyur}
% \email{{braden.webb, sumit.purohit, rounak.meyur}@pnnl.gov}
% \affiliation{%
%   \institution{Pacific Northwest National Laboratory}
%   \city{Richland}
%   \state{Washington}
%   \country{USA}
%   \postcode{99352}
% }

\maketitle

\submittednotice

\begin{abstract}
The integration of the Internet of Things (IoT) into Cyber-Physical Systems (CPSs) has expanded their cyber-attack surface, introducing new and sophisticated threats with potential to exploit emerging vulnerabilities.
Assessing the risks of CPSs is increasingly difficult due to incomplete and outdated cybersecurity knowledge. 
This highlights the urgent need for better-informed risk assessments and mitigation strategies. 
While previous efforts have relied on rule-based natural language processing (NLP) tools to map vulnerabilities, weaknesses, and attack patterns, recent advancements in Large Language Models (LLMs) present a unique opportunity to enhance cyber-attack knowledge completion through improved reasoning, inference, and summarization capabilities. 
We apply embedding models to encapsulate information on attack patterns and adversarial techniques, generating mappings between them using vector embeddings.
Additionally, we propose a Retrieval-Augmented Generation (RAG)-based approach that leverages pre-trained models to create structured mappings between different taxonomies of threat patterns.
Further, we use a small hand-labeled dataset to compare the proposed RAG-based approach to a baseline standard binary classification model.
Thus, the proposed approach provides a comprehensive framework to address the challenge of cyber-attack knowledge graph completion. 
\end{abstract}

\begin{IEEEkeywords}
Cybersecurity, Cyber-Physical Systems, Knowledge Graph, Retrieval Augmented Generation, Large Language Models
\end{IEEEkeywords}

\section{Introduction}\label{sec:intro}
% Overview of CAPECs and ATT&CK techniques
The integration of the Internet of Things (IoT) into Industrial Control Systems (ICS) has enhanced their automation, efficiency, and productivity in the industrial environment through a seamless convergence of the information technology (IT) and operational technology (OT) domains~\cite{ahmed2023}.
The widespread adoption of the Industrial Internet of Things (IIoT) has also created opportunities for cyber attacks, leading to the exploitation of the confidentiality, integrity, and availability (CIA) of the service and/or data~\cite{mekala2023}.
Some examples of such attacks include, but are not limited to, malware/ransomware attacks~\cite{ransom1,ransom2,ransom3}, distributed denial of service (DDoS)~\cite{ddos}, phishing attacks~\cite{phish}, and supply chain compromise~\cite{supply}.
Such threats can pose significant risk to critical infrastructure and have severe consequences for safety, economy, and public well-being~\cite{Mohammed2023}.

To understand how adversaries exploit vulnerabilities in cyber systems, the Common Attack Pattern Enumerations and Classifications (CAPEC), which are maintained by MITRE Corporation, serve as a publicly available catalog of cyber attack patterns\cite{capec}. 
Similarly, the MITRE Adversarial Tactics, Techniques, and Common Knowledge (ATT\&CK) framework is another essential tool that offers a comprehensive knowledge base for cyber threats from real world observations~\cite{mitre}. 
This framework provides a taxonomy of adversarial motivations or goals (\emph{tactics}) and a list of actions often taken to achieve those goals (\emph{techniques}) in Enterprise, mobile, and ICS networks.

These two publicly available information repositories, CAPEC and ATT\&CK, collectively contain a wealth of knowledge about cyber threats that could be invaluable to organizations seeking to model potential adversarial behavior, plan mitigation measures, or otherwise improve their cybersecurity infrastructure.
Therefore, we developed a novel approach and capability to better understand and model the relationship between the two taxonomies and identify the ATT\&CK techniques that correspond to each CAPEC attack pattern. 
While some cross-references currently exist for the Enterprise-oriented adversarial techniques in the MITRE ATT\&CK framework, there are no such references for techniques in the ICS or mobile domains.
A challenge to this integration is the lack of an automated framework which maps a CAPEC attack pattern to related ATT\&CK techniques.

% Large language models and domain specific usage

This integration is a daunting task. As of August 2024, there were 559 \capecs{} and 83 \icss{}, for a combination of 46,397 possible mappings between the two repositories. Moreover, significant domain expertise is necessary to verify whether a given connection is valid, often requiring discussion between cybersecurity experts, system operators, and users. Since they are frequently updated, it is difficult to justify manually linking the two data sets.

Although traditional machine learning classifiers might seem useful for identifying similar descriptions of attack patterns and techniques, most of these approaches require fairly structured input to learn representations.
Almost all of the information describing a \capec{} \ or \ics{}, however, is expressed in fields of unstructured text consisting of entry identifier, name and description. 
% We provide examples in Figure~\ref{fig:example_docs}.

We therefore turn to methodologies of natural language processing, where recent advances in large language models provide an opportunity to use artificial intelligence as a tool in automating the mapping task.
In particular, many embedding models can algorithmically encode text strings as arrays of floating-point numbers in a high-dimensional normed vector space. 
We can interpret these vectors, known as \textit{document embeddings} or simply \textit{embeddings}, as representing the semantics of the input text.
Embeddings of documents with similar meanings end up close together, and unrelated documents generate embeddings that are farther apart \cite{reimers-2019-sentence-bert}.
We use this process to treat difficult-to-handle unstructured text as mathematical vectors, to which we can then apply more standard machine learning tools. Specifically, we compare a mapping approach that identify nearest neighbors in embedding space with a retrieval-augmented generation (RAG) approach. We evaluate our results on a hand-labeled data set.

\noindent\textbf{Problem Statement.}~
% The main objective of this paper is to generate an automated framework required for the purpose of \emph{cyber attack knowledge graph completion}.
% A fundamental step in this framework is to evaluate a mapping between~\capecs and~\icss \ in order to integrate the two repositories.
% Further, we seek to enable a feature in the automated framework which adds newly identified mappings to the existing set every time a new entry (\capec \ or ICS ATT\&CK technique) is added to either repository.
% To this end, we use several large-scale language models to embed each~\capec \ and~\ics \ into high-dimensional vector spaces. 
% The hierarchical categories of~\capecs \ are used to compare the embedding models and identify the best option among them suitable for our purpose. 
% The resulting representations can then be processed using standard vector-based algorithms for evaluating relations between the~\capecs~and~\icss.
% Finally, we can use the same embedding model to generate mapping between~\capecs \ and~\icss.
In this paper, we address the key research challenge of \emph{cyber-attack knowledge graph completion}, specifically bridging gaps between disparate cybersecurity knowledge silos to support risk assessment and mitigation planning for Cyber-Physical Systems (CPSs). 
We focus on creating a bidirectional mapping between the CAPEC and ATT\&CK frameworks, ensuring that a \capec~and an \ics~are connected only when they accurately describe the same adversarial behavior. 
To achieve this, we use embedding models to encode text descriptions of \capecs{} and \icss{} into vectors, and use machine learning algorithms to generate the mappings. 
This paper explores the two following sub-problems: (i) evaluating various embedding models to determine the most effective one for \emph{cyber-attack knowledge graph completion}, and
(ii) generate and validate a mapping between \capecs{} and \icss{} using the vector embeddings obtained from their tokenized descriptions.
% For the first problem, we cluster \capecs into groups comprised of related attack patterns.
% We validate the clusters, which are identified after using the embedding models, against the \textit{ground-truth} hierarchical categories listed in~\cite{capec}.
% This allows us to compare the embedding models and identify the one best suited for the task at hand.
% Hereafter, we use this model to evaluate representations of \capecs{} and \icss{} in the embedding space.
% We use the proximity of these representations in the embedding space to address our second problem.

\noindent\textbf{Contributions.}~
% \noindent\textbf{Contributions.}~
% Our contributions in this paper are as follows: 
% (i) we compare state-of-the-art embedding models for the cyber-attack knowledge graph completion task,
% identify and apply suitable metrics for evaluating proposed 
% both partitioning the CAPEC and MITRE ATT\&CK databases as well as for classifying their relationships,
% (ii) we demonstrate the viability of both standard classification methods and a RAG-based approach to generate mappings, and
% (iii) we publish a small hand-labeled dataset that captures relationships between \capec{} and \ics{}, providing a critical resource for validating the proposed methodology.
% (i) we define suitable metrics to evaluate the performance of several language models used to map \capecs to the \ics, 
% (ii) identify key roadblocks while using off-the-shelf language models in the context of cybersecurity, 
% (iii) 
This paper makes several key contributions to the field of cyber-attack knowledge graph completion. 
First, we provide a comprehensive comparison of state-of-the-art embedding models when used for the cyber-attack knowledge graph completion task. 
Second, we demonstrate the effectiveness of both traditional classification methods and a RAG-based approach in generating accurate mappings between cybersecurity taxonomies. 
Finally, we contribute a valuable resource to the community by publishing a small, hand-labeled dataset that captures the relationships between \capecs{} and \icss{}, which serves as a critical tool for validating our proposed methodology.

The remainder of the paper is outlined as follows: section \ref{sec:rel} gives an overview of related work and section \ref{sec:method} presents embedding and mapping generation process. We discuss evaluation metrics in section \ref{sec:eval} and the results in section \ref{sec:res}.
\section{Related Work}\label{sec:rel}

% on the MITRE databases

% on language modeling
Researchers have explored AI/ML approaches to automate the process of cyber knowledge alignment across different data sources. Random Forest \cite{aota2020automation}, naive bayes classifier \cite{na2017study}, and natural language-based similarity measures \cite{kanakogi2021tracing} has been used with limited success to align several MITRE repositories, including Common Vulnerabilities and Exposures (CVE), Common Weakness Enumeration (CWE), and CAPEC. Maunero et al. \cite{maunero2023towards} use an ontology-based approach to automate the risk assessment process. These approaches suffer from a lack of ground-truth data required for model training and rule generation. In contrast, our previous work \emph{V2W-BERT} ~\cite{sumit2022} and Villanueva-Miranda et al. \cite{villanueva2023analyzing} have also demonstrated the use of encoder-only and encoder-decoder language models, such as BERT and Google T5, to automate the task of mapping between various cybersecurity databases with high accuracy. However, decoder-only large language models have greatly improved in recent years and have revolutionized the way they are used in \textit{few-shot} learning applications with limited data. In this paper, we use state-of-the-art embedding models for this purpose and performed a comparison among them to identify which among them suits best for the task of cyber knowledge completion, and utilize a decoder-only LLM to refine their output.

\section{Methods} \label{sec:method}
This section introduces mathematical notations used to describe the cyber-knowledge problem space and presents an embedding and mapping generation approach. We provide examples of the descriptive text contained within \capecs{} and \icss{}, and use it to explain the methodology.  
\subsection{Preliminaries}
Let $C = \{c_1, \dots, c_{N}\}$ be the set of $N$ CAPEC attack patterns and $T=\{t_1, \dots, t_{M}\}$ the set of $M$ MITRE ATT\&CK ICS techniques. We want to find the set of mappings $\mathcal{M} \subseteq C \times T$ such that for each $(c, t) \in \mathcal{M}$, adversarial behaviors can be accurately described by both $c$ and $t$. We can represent this set by a function $f : C \to \mathcal{P}(T)$ from individual \capecs{} to sets of techniques, where $\mathcal{P}(T)$ denotes the power set of $T$.
A possible approximation for this function is to identify the best $k$ choices of ICS techniques that can be related to a given CAPEC attack pattern, i.e., 
$f_k(c_i) = \{t_{i1}, \dots, t_{ik}\}$, where $t_{i1}, \dots, t_{ik}$ are the $k$ ICS techniques that are most similar to $c_i$. 

We can also model the problem in the reverse direction as the process of learning a function $g : T \to \mathcal{P}(C)$. Indeed, our methods work in both directions (although the asymmetry produces slightly different results). However, since the process is completely analogous, we focus on describing the ``forward`` direction $C \to \mathcal{P}(T)$ in the remainder of the paper.
% The value of $k$ differs for each $c_i$; i.e., $k$ is a function from $C$ to $\{1, \dots, M\}$. We can equivalently think of this function $f$ in terms of it's set-theoretic definition: $f \subseteq C \times T$. In these terms, a mapping $(c, t) \in C \times T$ is an association of one CAPEC attack pattern to an ICS technique, and $(c, t) \in f$ if and only if the two are semantically related.

\begin{figure}
  \centering
  \small
    \begin{center}
      \textbf{A CAPEC Attack Pattern}
  \end{center}
  \begin{tabular}{m{\columnwidth}}
    \textbf{ID:} CAPEC-125 \\
    \textbf{Name:} Flooding \\
    \textbf{Description}: An adversary consumes the resources of a target by rapidly engaging in a large number of interactions with the target. This type of attack generally exposes a weakness in rate limiting or flow. When successful this attack prevents legitimate users from accessing the service and can cause the target to crash. This attack differs from resource depletion through leaks or allocations in that the latter attacks do not rely on the volume of requests made to the target but instead focus on manipulation of the target's operations. The key factor in a flooding attack is the number of requests the adversary can make in a given period of time. The greater this number, the more likely an attack is to succeed against a given target.
  \end{tabular} 
  \\
  \begin{center}
      \textbf{A MITRE ICS ATT\&CK Technique}
  \end{center}
  \begin{tabular}{m{\columnwidth}}
    \textbf{ID:} T0814 \\
    \textbf{Name:} Denial of Service \\
    \textbf{Description:} Adversaries may perform Denial-of-Service (DoS) attacks to disrupt expected device functionality. Examples of DoS attacks include overwhelming the target device with a high volume of requests in a short time period and sending the target device a request it does not know how to handle. Disrupting device state may temporarily render it unresponsive, possibly lasting until a reboot can occur. When placed in this state, devices may be unable to send and receive requests, and may not perform expected response functions in reaction to other events in the environment. Some ICS devices are particularly sensitive to DoS events, and may become unresponsive in reaction to even a simple ping sweep. Adversaries may also attempt to execute a Permanent Denial-of-Service (PDoS) against certain devices, such as in the case of the BrickerBot malware.
  \end{tabular} 
  \caption{Examples of \emph{description strings} for a \capec{} \cite{capec} and an \ics{} \cite{mitre} which describe very similar adversarial behavior. A good framework should generate a mapping between these two documents.}
  \label{fig:example_docs}
\end{figure}

\subsection{Embedding Generation}
Our approach to learn the mapping function between the CAPEC and ATT\&CK taxonomies utilizes document embedding models. 
Specifically, we use transformer-based neural networks that produce fixed-length, dense representations of variable length documents---allowing us to compare the taxonomies quantitatively. 

\begin{table}[!t]
    \centering
    \caption{The embedding models that we compared and evaluated, and the fixed size $d$ of the embeddings they generate.}
    \begin{tabular}{l|c}
        Embedding Model $\Phi$ & Dimensionality $d$ \\
        \hline
        \texttt{text-embedding-ada-002} \cite{ada-002} & 1536 \\
        \texttt{E5-large-v2} \cite{wang2022text} & 1024 \\
        \texttt{instructor-large} \cite{INSTRUCTOR} & 768 \\
        \texttt{all-MiniLM-L6-v2} \cite{reimers-2019-sentence-bert} & 384 
    \end{tabular}
    \label{tab:embed_models}
\end{table}

% \begin{table}[!t]
%     \centering
%     \caption{Possible $\Omega$ Functions}
%     \label{table_omega}
%     \begin{IEEEeqnarraybox}[\IEEEeqnarraystrutmode\IEEEeqnarraystrutsizeadd{2pt}{1pt}]{v/c/v/c/v}
%         \IEEEeqnarrayrulerow\\
%         &\mbox{Range}&&\Omega(m)&\\
%         \IEEEeqnarraydblrulerow\\
%         \IEEEeqnarrayseprow[3pt]\\
%         &x < 0&&\Omega(m)=\sum\limits_{i=0}^{m}K^{-i}&\IEEEeqnarraystrutsize{0pt}{0pt}\\
%         \IEEEeqnarrayseprow[3pt]\\
%         \IEEEeqnarrayrulerow\\
%         \IEEEeqnarrayseprow[3pt]\\
%         &x \ge 0&&\Omega(m)=\sqrt{m}\hfill&\IEEEeqnarraystrutsize{0pt}{0pt}\\
%         \IEEEeqnarrayseprow[3pt]\\
%         \IEEEeqnarrayrulerow
%     \end{IEEEeqnarraybox}
% \end{table}

We prepare a \emph{description string} for each \capec{} and \ics{} by concatenating their name, ID, and description, as displayed in Fig. \ref{fig:example_docs}. An embedding model, $\Phi$, then tokenizes each input \emph{description string} to a list of tokens and transforms those tokens to a vector in $\mathbb{R}^d$. Although most of the \emph{description strings} are quite short and fit within the maximum sequence length of the models, truncation is performed for the few longer descriptions where necessary. Our specific embedding models and their corresponding dimensions are listed in Table \ref{tab:embed_models}.

\subsection{Mapping Generation}
The next task is to use the vector embedding of the \emph{description strings} of \capecs{} and \icss{} to generate a mapping between the taxonomies, i.e., identify the list of $k$ entries from one taxonomy that are most similar to a given entry from the other taxonomy.
Here, we present two methods of generating the required mapping -- (i) nearest-neighbour mapping, and (ii) RAG-based mapping. These are compared in Fig. \ref{fig:rag_flowchart}. As we will show in this section, the RAG-based approach builds directly off the nearest-neighbor approach, refining its output to improve precision.

\subsubsection*{Nearest-Neighbor Mapping}\label{sec:nnmap}
% \sloppy\noindent\textbf{Nearest-Neighbor Mapping}
\sloppy Treating the direction of the vector embedding of the \emph{description string} as indicative of its semantic meaning, we can approach our problem of identifying $f_k(c_i) = \{t_{i1}, \dots, t_{ik}\}$ by evaluating the $k$-nearest neighbors of the CAPEC embedding $\Phi\left(c_i\right)$ from the set of technique embeddings $\left\{\Phi\left(t_1\right),\Phi\left(t_2\right),\cdots,\Phi\left(t_M\right)\right\}$. Formally, then,

\[f_{k, \Phi}(c_i) = \argmax_{J \subset T: |J| = k } \sum_{t \in J} \Phi(c_i)^T\Phi(t),\] 
which is the set of $k$ techniques that have the largest inner products with $\Phi(c_i)$. At this scale, matrix multiplication makes this calculation easily tractable. As a result, for each choice of both embedding model and $k$, we have complete approximate mappings given by 
\[
    \underset{c \in C}{\Bigl\{\bigl(c, t \bigr) : t \in f_{k, \Phi}(c)\Bigr\}} 
    \quad \text{and} 
    \quad \underset{t \in T}{\Bigl\{\bigl(c, t \bigr): c \in g_{k, \Phi}(t)\Bigr\}}
\]

% \[
% \Bigl\{ \bigl(c, t\bigr) : c \in C, t \in f_{k, \Phi}(c) \Bigr\} 
% \quad \text{and} \quad 
% \Bigl\{ \bigl(c, t\bigr) : t \in T, c \in g_{k, \Phi}(t) \Bigr\}
% \]

% \[
%     \bigcup_{c \in C}{\Bigl\{\bigl(c, t \bigr) : t \in f_{k, \Phi}(c)\Bigr\}} 
%     \quad \text{and} \quad 
%     \bigcup_{t \in T}{\Bigl\{\bigl(c, t \bigr): c \in g_{k, \Phi}(t)\Bigr\}}
% \]

for both the forwards (CAPEC to ATT\&CK) and backwards (ATT\&CK to CAPEC) directions. For notational convenience going forward, we will not write $\Phi(c)$ or $\Phi(t)$ each time we refer to the embedding of a \capec{} or \ics{}. Instead we also use $c$ or $t$ to denote their embedded representation in $\mathbb{R}^d$.

% There are $|\mathcal{P}(T)|^{|C|} = (2^{93})^{615} = 2^{57195}$ possible functions. 

% There is also the question of which pairs, $(c, t)$ are valid relationships. The set of all such pairs is the Cartesian product, $C \times T$, with cardinality 
% \[|C \times T| = |C| \times |T| = 615 \times 93 = 57195.\]

% It is not a coincidence that there are $|\mathcal{P}(C \times T)| = 2^{615\times 93}$ possible combinations of these ordered pairs.

% So although there are only $57195$ items that we're trying to identify, there are $2^{57195}$ possible answers to the entire question (i.e. elements of the function space).

% \input{clustering}

The nearest-neighbor embedding approach provides a baseline method of retrieving potential candidates for CAPEC-ATT\&CK mappings, but suffers from filtering those candidates with precision. It also requires every \capec{} (or \ics{}) to be linked to the same fixed number $k$ of \icss{} (or \capecs{}), when in reality the number of links might vary widely.
Hence, we present a RAG-based mapping as an alternative to address these common problems.

\subsubsection*{RAG-Based Mapping}

\begin{figure*}[ht]
    \centering
    \includegraphics[scale=.55]{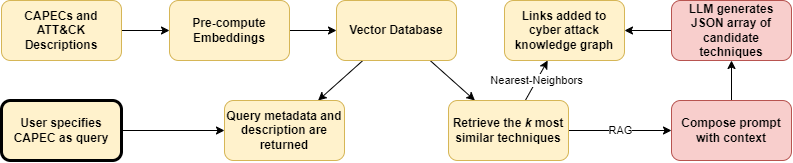}
    \caption{The nearest-neighbor and RAG pipelines for cyber attack knowledge graph completion, shown in the CAPEC-to-ATT\&CK direction. Modules in yellow are common to both the nearest-neighbor and RAG pipelines, while those in red are unique to the RAG-based approach.}
    \label{fig:rag_flowchart}
\end{figure*}

We propose a method that improves upon the simple embedding-based retrieval method by utilizing LLMs in an approach similar to standard RAG techniques~\cite{lewis_perez_rag}. This RAG pipeline relies upon both an embedding model and a generative language model---the nearest-neighbor mapping function $f_{k, \Phi}$ is in fact the first step of the RAG pipeline. As shown in Fig. \ref{fig:rag_flowchart}, an individual \capec{} $c$ is fed into the pipeline as input along with a parameter $k$, and the resulting techniques $f_{k, \Phi}(c)$ are retrieved. While we did store the intermediate, low-dimensional embedded representations in a vector database, the taxonomies are currently small enough that all embeddings can alternatively be kept in memory. Once retrieved, the techniques are ranked according to their proximity to $c$, and then passed along to the LLM in a prompt. Because of the significant instability and sensitivity of LLMs to small changes in their inputs \cite{wang2024prompt}, prompt engineering is necessary to improve their robustness. We utilized the open-source 8B-parameter instruction fine-tuned variant of Meta's Llama 3 for this purpose \cite{llama3modelcard}. 

Because a key desideratum of our research is to automate this mapping task, we also desired the outputs to be structured in a predictable, machine-accessible format. To this end, we leverage a decoding technique to sample generated tokens according to any context-free grammar. This allows us to specify a JSON schema of our choice that we can then convert to a Backus-Naur form of a formal grammar \cite{mccracken2003backus} to constrain the LLM's output tokens accordingly \cite{ggerganovLlamacpp}. 

 The retrieved information is added as a context to the LLM. In contrast to the nearest-neighbor mapping presented earlier, the RAG-based approach leverages the LLM's summarizing and reasoning capabilities to provide explainable mapping results.

\section{Evaluation}\label{sec:eval}

The lack of any external ground truth data set for validating our results is a major roadblock in efforts to evaluate the efficacy of our methodology. 
% The below passage was copied in from later in this same section, because I (Braden) believe it flows much better right here.
Indeed, this data-scarcity problem is a key aspect of the very problem we seek to address---and this is reflected in our test set. We hand-crafted this sample by manually labeling a set of what we determined to be corresponding \capecs{} and \icss{}. While we sought to obtain a representative sample of both taxonomies, the selection process was not entirely random. Several \capecs{} were chosen because they were either (i) classified as being `meta'-level abstractions or (ii) listed by MITRE among patterns that could be relevant to industrial control systems. However, it is quite possible that there are other, equally relevant \icss{} for any one of those \capecs{}. Moreover, it is highly unlikely that each of the \capecs{} chosen to generate the test set should map to the same number $k$ of \icss{} (and indeed, for a given $c \in C$, the size of the set given by $\{t : (c, t) \in G\}$ varies between 0 and 5). For these reasons, we explicitly tried to get as broad of a range of patterns and techniques included in the data set as possible and developed novel metrics to gain better insight into pipeline performance.

Due to the many-to-many relationship that should exist between the CAPEC database and the MITRE ATT\&CK framework, we determined to evaluate our approach as a standard classification model. In general, important performance metrics for a classification algorithm include:
\begin{itemize}
    \item \textbf{accuracy} --- the ratio of correct classifications (both positive and negative) to all possible classifications
    \item \textbf{recall} --- the ratio of correctly retrieved relevant instances (true positives) to all relevant instances
    \item \textbf{precision} --- the ratio of correctly retrieved relevant instances (true positives) to all retrieved instances
    \item \textbf{$F_1$-score} --- the harmonic mean of precision and recall, given by
    \[F_1 = 2 \times \frac{\text{precision} \times \text{recall}}{\text{precision} + \text{recall}}\]
\end{itemize}
In order to make inferences about a larger population, or for reasons of computational efficiency, these metrics are usually calculated on a test set of randomly sampled data points rather than on the entire space of possibilities.

We provide these same metrics in the language of our problem, but due to insufficiencies in our test set, we also define notions of \textit{coverage} and a \textit{false mapping ratio}. Since many of the methods we discuss involve computing various 1-to-$k$ mappings from CAPEC attack patterns to ICS techniques, these metrics allow us greater insight into how often, on average, a given function maps individual attack patterns to at least one of their associated techniques, as well as how often we should expect false positives among our retrieved mappings. The metrics can also be reversed to apply analogously for mappings from ATT\&CK space to CAPEC space.

\subsection{Metric Definitions}
% Are these set-theoretic definitions of recall and precision sufficient if we don't have negative examples in the test set as well?

Let $C$ be the set of all CAPEC attack patterns and $T$ the set of all MITRE ICS techniques. We consider  mappings $\mathcal{M},~G~\subseteq~C\times~T$, where $\mathcal{M}$ is the mapping we wish to evaluate, and $G$ is the labeled set of ground-truth mappings. In other words, $G$ denotes a mapping such that for each pair $(c, t) \in G$, the CAPEC attack pattern $c$ truly does correspond to $t$. We can then define 

\[\text{recall}_G(\mathcal{M}) = \frac{|\mathcal{M} \cap G|}{|G|}\]

and

\[\text{precision}_G(\mathcal{M}) = \frac{|\mathcal{M} \cap G|}{|\{(c_i, t_i) \in \mathcal{M} : \exists t \in T, (c_i, t) \in G\}|}.\]
In essence, when calculating the precision of $\mathcal{M}$ given $G$, we only want to evaluate $\mathcal{M}$ over the CAPEC attack patterns considered in $G$.

It is therefore natural to define the $F$-score (or $F_1$-score) of $\mathcal{M}$ given $G$ as

\[F_G(\mathcal{M}) = 2 \times \frac{\text{precision}_G(\mathcal{M}) \times \text{recall}_G(\mathcal{M})}{\text{precision}_G(\mathcal{M}) + \text{recall}_G(\mathcal{M})}.\]

% \rounak{`Recall' is the fraction of relevant objects that have been retrieved. In our case, the \emph{relevant} objects are enlisted in the set $G$ of ground truth mappings. The \emph{retrieved} objects are the set of mappings $\mathcal{M}$, which have been identified by the proposed methodology.}

% \rounak{However, the proposed embedding model may not be able to retrieve all mappings in $G$. Hence, we relax the definition of `recall' as follows: the fraction of \capecs which has been mapped to at least one relevant \ics.}

%%%%%%%%%%%%%%%%%%%%%%%
% We had a short section here that I moved to the top of the Evaluation section instead
%%%%%%%%%%%%%%%%%%%%%%%

\noindent\textbf{Coverage.}~In this context, we find it useful to introduce a notion of \textit{coverage}. In doing so, the number of relevant pairs are computed for each $c\in C_G = \{c \in C : \exists t \in T, (c, t) \in G\}$, and we define $\mathcal{M}_c\subseteq \mathcal{M}$ as the set of all maps from attack pattern $c$ to one or more ICS techniques:
\[\mathcal{M}_c = \left\{ \left(c_i,t_i\right) \in \mathcal{M}, c_i=c \right\} \]

We then define the coverage, with respect to $G$, of a mapping $\mathcal{M}$ to be the proportion of attack patterns $c$ in $C_g$ for which $\mathcal{M}_c$ contains a valid mapping from $G$, i.e.,

\begin{equation*}
    \text{coverage}_G(\mathcal{M}) = \frac{\sum_{c\in C_G}{\mathbb{I}\left[\mathcal{M}_{c} \bigcap G \neq \emptyset \right]}}{|C_G|}
\end{equation*}

While we could just as easily use a notion of technique coverage rather than CAPEC coverage, this method is more easily interpretable, given our test set, as the proportion of \capecs{} considered in $\mathcal{M}$ that are mapped to at least one of the techniques identified in $G$.

\noindent\textbf{False Mapping Ratio (FMR).}~
We also propose a metric, which we call the \gls{fmr}, to evaluate the frequency with which our predicted mapping methodology selects non-similar pairs. 
The \gls{fmr} is defined as the fraction of element pairs that are explicitly labeled as non-similar in the ground truth dataset but are erroneously identified as similar by the proposed mapping algorithm. 
Mathematically, the number of false positive mappings (non-similar pairs predicted as similar) can be denoted by \(\mathcal{M} \bigcap \widetilde{G}\), where \(\widetilde{G}\) is the total number of non-similar pairs explicitly identified in the ground truth dataset. The \gls{fmr} is then expressed as:

\[
\text{FMR} = \frac{|\mathcal{M} \bigcap \widetilde{G}|}{|\widetilde{G}|}
\]

A higher \gls{fmr} indicates poorer performance, as it shows the proposed model’s tendency to incorrectly map non-similar elements. Conversely, a lower FMR suggests better mapping accuracy.

\section{Results}\label{sec:res}

\begin{table*}[ht]
    \centering
    \caption{CAPEC-to-ATT\&CK Results}
    \label{tab:results-c2t}
    \begin{tabular}{cc|lllll|lllll}
    \toprule
    &  & \multicolumn{5}{c}{Nearest neighbor mapping} & \multicolumn{5}{c}{RAG-based mapping} \\
    Model & $1$-to-$k$ & Recall $\uparrow$ & Precision $\uparrow$ & F-Score $\uparrow$ & Coverage $\uparrow$ & \gls{fmr} $\downarrow$ & Recall $\uparrow$ & Precision $\uparrow$ & F-Score $\uparrow$ & Coverage $\uparrow$ & \gls{fmr} $\downarrow$ \\ \midrule
\multirow{5}{*}{e5}
        & 1-to-1 & .0700 & .3684 & .1176 & .3684 & .0833 & .0769 & .2857 & .1212 & .4000 & .1077  \\
        & 1-to-2 & .1700 & .4474 & .2464 & .5789 & .1389 & .1400 & .4667 & .2154 & .5263 & .1111  \\
        & 1-to-3 & .2200 & .3860 & .2803 & .5789 & .2407 & .1500 & .4412 & .2239 & .4737 & .1481  \\
        & 1-to-4 & .3300 & .4342 & .3750 & .8421 & .2778 & .2020 & .5263 & .2920 & .6667 & .1596  \\
        & 1-to-5 & .4100 & .4316 & .4205 & .8947 & .3426 & .2300 & .5000 & .3151 & .6316 & .1574  \\
        \midrule
\multirow{5}{*}{instructor}
        & 1-to-1 & .1000 & .5263 & .1681 & .5263 & .0648 & .0920 & .5333 & .1569 & .5333 & .0722  \\
        & 1-to-2 & .1900 & .5000 & .2754 & .7368 & .1389 & .1771 & .6071 & .2742 & .7778 & .0943  \\
        & 1-to-3 & .2700 & .4737 & .3439 & .7895 & .2037 & .2000 & .5714 & .2963 & .7368 & .1111  \\
        & 1-to-4 & .3500 & .4605 & .3977 & .8947 & .2778 & .2386 & .5833 & .3387 & .7647 & .1429  \\
        & 1-to-5 & \textbf{.4200} & .4421 & \textbf{.4308} & .8947 & .3148 & .2200 & .5116 & .3077 & .6842 & .1481  \\
        \midrule
\multirow{5}{*}{sent-transf}
        & 1-to-1 & .1000 & .5263 & .1681 & .5263 & .0741 & .0610 & .3846 & .1053 & .3846 & .0959  \\
        & 1-to-2 & .1900 & .5000 & .2754 & .7368 & .1389 & .1505 & .5833 & .2393 & .7500 & .1034  \\
        & 1-to-3 & .2400 & .4211 & .3057 & .7368 & .2315 & .1735 & .5484 & .2636 & .7778 & .1313  \\
        & 1-to-4 & .3200 & .4211 & .3636 & \textbf{.9474} & .2870 & .2143 & .5122 & .3022 & .8333 & .1717  \\
        & 1-to-5 & .3800 & .4000 & .3897 & \textbf{.9474} & .3519 & \textbf{.2449} & .4706 & .3221 & \textbf{.8889} & .2020  \\
        \midrule
\multirow{5}{*}{ada-002}
        & 1-to-1 & .1200 & \textbf{.6316} & .2017 & .6316 & \textbf{.0556} & .1084 & \textbf{.6429} & .1856 & .6429 & \textbf{.0610}  \\
        & 1-to-2 & .1900 & .5000 & .2754 & .6842 & .1019 & .1856 & .6000 & .2835 & .7647 & .0941  \\
        & 1-to-3 & .2400 & .4211 & .3057 & .7895 & .1667 & .1900 & .5758 & .2857 & .7368 & .1019  \\
        & 1-to-4 & .3000 & .3947 & .3409 & .8947 & .2130 & .2200 & .5641 & .3165 & .7895 & .1204  \\
        & 1-to-5 & .3300 & .3474 & .3385 & \textbf{.9474} & .2407 & .2400 & .6000 & \textbf{.3429} & .7895 & .1296  \\
\bottomrule
    \end{tabular}
\end{table*}

\begin{table*}[ht]
    \centering
    \caption{ATT\&CK-to-CAPEC Results}
    \label{tab:results-t2c}
    \begin{tabular}{cc|lllll|lllll}
    \toprule
    &  & \multicolumn{5}{c}{Nearest neighbor mapping} & \multicolumn{5}{c}{RAG-based mapping} \\
    Model & $1$-to-$k$ & Recall $\uparrow$ & Precision $\uparrow$ & F-Score $\uparrow$ & Coverage $\uparrow$ & \gls{fmr} $\downarrow$ & Recall $\uparrow$ & Precision $\uparrow$ & F-Score $\uparrow$ & Coverage $\uparrow$ & \gls{fmr} $\downarrow$ \\ \midrule
\multirow{5}{*}{e5}
        & 1-to-1 & .0769 & .6000 & .1364 & .6000 & .0145 & .0882 & .6667 & .1558 & .6667 & .0114  \\
        & 1-to-2 & .1282 & .5000 & .2041 & .8000 & .0364 & .1282 & .5882 & .2105 & .8000 & .0255  \\
        & 1-to-3 & .1923 & .5000 & .2778 & .7000 & .0545 & .1765 & .7500 & .2857 & .7778 & .0152  \\
        & 1-to-4 & .2692 & .5250 & .3559 & .8000 & .0691 & .2051 & .7273 & .3200 & .7000 & .0218  \\
        & 1-to-5 & .3333 & .5200 & .4062 & .8000 & .0873 & \textbf{.2941} & .8000 & .4301 & .8889 & .0189  \\
        \midrule
\multirow{5}{*}{instructor}
        & 1-to-1 & .0641 & .5000 & .1136 & .5000 & .0182 & .0641 & .5000 & .1136 & .5000 & .0182  \\
        & 1-to-2 & .1410 & .5500 & .2245 & .7000 & .0327 & .1471 & .6250 & .2381 & .7778 & .0227  \\
        & 1-to-3 & .2308 & .6000 & .3333 & .8000 & .0436 & .1667 & .6500 & .2653 & .7000 & .0255  \\
        & 1-to-4 & .2949 & .5750 & .3898 & .8000 & .0618 & .2308 & .7500 & .3529 & .8000 & .0218  \\
        & 1-to-5 & .3590 & .5600 & .4375 & .8000 & .0800 & .2436 & .7600 & .3689 & .8000 & .0218  \\
        \midrule
\multirow{5}{*}{sent-transf}
        & 1-to-1 & .1026 & .8000 & .1818 & .8000 & .0073 & .1026 & .8000 & .1818 & .8000 & .0073  \\
        & 1-to-2 & .2051 & .8000 & .3265 & .9000 & .0145 & .1923 & .8824 & .3158 & .9000 & .0073  \\
        & 1-to-3 & .2692 & .7000 & .3889 & .9000 & .0327 & .2051 & .8421 & .3299 & .9000 & .0109  \\
        & 1-to-4 & .3333 & .6500 & .4407 & .9000 & .0509 & .2436 & .7600 & .3689 & .8000 & .0218  \\
        & 1-to-5 & .3718 & .5800 & .4531 & .9000 & .0764 & .2564 & .7692 & .3846 & .9000 & .0218  \\
        \midrule
\multirow{5}{*}{ada-002}
        & 1-to-1 & .1235 & \textbf{.8333} & .2151 & .8333 & \textbf{.0070} & .1235 & .8333 & .2151 & .8333 & \textbf{.0070}  \\
        & 1-to-2 & .1923 & .7500 & .3061 & \textbf{1.0000} & .0182 & .1667 & .8667 & .2796 & \textbf{1.0000} & .0073  \\
        & 1-to-3 & .2564 & .6667 & .3704 & \textbf{1.0000} & .0364 & .1795 & .7778 & .2917 & \textbf{1.0000} & .0145  \\
        & 1-to-4 & .3333 & .6500 & .4407 & \textbf{1.0000} & .0509 & .2564 & .8696 & .3960 & \textbf{1.0000} & .0109  \\
        & 1-to-5 & \textbf{.4103} & .6400 & \textbf{.5000} & \textbf{1.0000} & .0655 & .2821 & \textbf{.9167} & \textbf{.4314} & \textbf{1.0000} & .0073  \\
\bottomrule
    \end{tabular}
\end{table*}

Table~\ref{tab:results-c2t} compares the two proposed mapping methodologies---nearest neighbor mapping and RAG-based mapping---from \capecs{} to \icss{}, while Table~\ref{tab:results-t2c} compares performance from ATT\&CK to CAPEC.
We compare four embedding models  for each mapping: \mbox{\textit{E5-large-v2}} (listed as `e5' in the tables) \cite{wang2022text}, \mbox{\textit{instructor-large}} (listed as `instructor') \cite{INSTRUCTOR}, \mbox{\textit{all-MiniLM-L6-v2}} (listed as `sent-transf') \cite{reimers-2019-sentence-bert}, and \mbox{\textit{text-embedding-ada-002}} (listed as `ada-002') \cite{ada-002}. 
The models are evaluated for mapping performance using standard metrics like Recall, Precision, and F-Score, as well as the new metrics introduced in section \ref{sec:eval}, namely Coverage and the \gls{fmr}. 
The best results are highlighted in bold.
By placing the symbols $\uparrow$ and $\downarrow$ next to the metric names, we indicate the direction in which an improvement is represented.
We consider $1$-$5$ nearest mapped ATT\&CK techniques for each CAPEC in order to evaluate the mapping between elements of the two frameworks.

The key findings for the mapping from \capecs{} to \icss{} shown in Table~\ref{tab:results-c2t} can be listed as follows: (i) RAG-based mapping generally outperforms nearest neighbor mapping in terms of precision and F-score across most embedding models, indicating more accurate mapping predictions. 
(ii) Coverage increases consistently as we move from 1-to-1 to 1-to-5 mappings for both methods. 
This means that considering more nearest neighbors allows for more elements to be mapped, though it may also increase false positives.
This can be validated from the \gls{fmr} scores, where we see that for each embedding model, the \gls{fmr} tends to increase as $k$ increases.
Recall that a lower \gls{fmr} indicates fewer incorrect mappings.
(iii) \textit{instructor-large} and \textit{text-embedding-ada-002} exhibit stronger performance in precision and F-score in both mapping methodologies, particularly in the RAG-based approach. 
\textit{E5-large-v2} performs consistently weaker compared to other models, with lower recall and F-scores in both methodologies.
Therefore, \textit{text-embedding-ada-002} and \textit{instructor-large} should be preferred as top choices of embedding model for this task.

The key findings for the mapping from ATT\&CK to CAPEC shown in Table~\ref{tab:results-t2c} can be listed as follows: 
(i) The performance generally improves as the number of nearest neighbors \(k\) increases from $1$ to $5$ for all embedding models and mapping methodologies.
(ii) The \gls{fmr} tends to increase with the number of neighbors \(k\), indicating a trade-off between coverage and false-positive matches. We note this holds only for the nearest neighbor mapping method.
(iii) The RAG-based mapping typically outperforms thenearest neighbor mapping across most metrics.
We note higher precision, coverage and F-scores for high \(k\) values, while the \gls{fmr} values are lower, indicating the superiority of RAG-based mapping for this task.
(iv) The \textit{text-embedding-ada-002} model shows the highest coverage, reaching $100\%$ for \(k \geq 2\) with both mapping methods, denoting it to be the best choice of embedding model for this mapping task.

A significant challenge in evaluating mapping methodologies between data sets, such as \capecs{} and \icss{}, is the lack of labeled ground truth dataset. 
Without a comprehensive, annotated mapping of relationships between these frameworks, it is difficult to assess the true accuracy of various predictive mapping models. 
Community-driven efforts are essential to address this gap by creating and maintaining a labeled data set that defines the relationships between CAPEC and ATT\&CK.
Such initiatives would not only enrich the cybersecurity knowledge graph but also provide a critical resource for data-driven approaches aimed at automating the mapping process. 
The availability of a well-labeled ground truth would enable researchers and practitioners to validate new methodologies, refine existing models, and improve the overall accuracy of threat detection and defense strategies within the cybersecurity landscape.

\section{Conclusion}
This study presents a comprehensive evaluation of mapping methodologies between two distinct taxonomies, leveraging both nearest neighbor and RAG-based approaches across multiple embedding models.
The results consistently demonstrate that RAG-based mapping outperforms nearest neighbor mapping in terms of precision, F-score, and the ability to reduce incorrect mappings, as evidenced by lower \gls{fmr} scores.
Among the embedding models, \textit{instructor-large} and \textit{text-embedding-ada-002} achieve the highest mapping accuracy, particularly when larger sets of neighbors are considered (1-to-5 mappings). 
Conversely, the \textit{E5-large-v2} embedding model consistently underperforms across both methodologies. 
These findings highlight the importance of selecting both an appropriate mapping strategy and embedding model when tackling the similarity-based mapping problem between elements of the CAPEC and ATT\&CK frameworks.

Future work will focus on refining these approaches through fine-tuning the LLMs to further reduce false mappings and improve scalability. 
Additionally, incorporating more advanced validation techniques, including the use of \textit{expert-in-the-loop} systems, could enhance the reliability and interpretability of the mappings.
% We will expand upon the \textit{ground-truth} dataset and develop an \textit{expert-in-the-loop} evaluation pipeline. 
We will also predict mappings between the CWE and CVE knowledge sources to provide a comprehensive cybersecurity risk assessment. 
% \begin{itemize}
%     \item uncertainty quantification~\cite{Xiao2022UncertaintyQW}
% \end{itemize}

\section*{Acknowledgement}
The research described in this paper is part of the Resilience Through Data Driven, Intelligently Designed Control (RD2C) Initiative at Pacific Northwest National Laboratory (PNNL). It was conducted under the Laboratory Directed Research and Development Program at PNNL, a multiprogram national laboratory operated by Battelle for the U.S. Department of Energy.

% \bibliography{assets/references}
% \bibliographystyle{plainnat}

\bibliographystyle{IEEEtran}
\bibliography{IEEEabrv,assets/references}

\end{document}